# SoilGen: A Comprehensive Tool for Generating Synthetic Soil Profiles for Geotechnical and Seismic Analysis

## Mersad Fathizadeh[1], Hosna Kianfar[2]

[1]University of Arkansas, Graduate Research Assistant, Dept. of Civil Eng., 4190 Bell Engineering Center Fayetteville, AR 72701, USA, mersadf@uark.edu
[2]University of Arkansas, Graduate Research Assistant, Dept. of Civil Eng., 4190 Bell Engineering Center Fayetteville, AR 72701, USA, hkianfar@uark.edu

**ABSTRACT**

Geotechnical and seismic applications, ranging from site response analysis and HVSR simulations to dispersion curve modeling, increasingly depend on large, well-labeled datasets for robust model development. However, the scarcity of publicly available borehole datasets—coupled with the proprietary nature of high-quality field records—creates a significant bottleneck for data-driven research, particularly in machine learning. To address this limitation, this study introduces SoilGen, an open-source framework that procedurally generates physically consistent, multilayered soil columns as synthetic soil profiles. Unlike simple randomization, SoilGen computes a complete suite of geotechnical properties—including thickness, $V_S$, P-wave velocity ($V_P$, Density and Poisson's ratio—while enforcing physical constraints to ensure realism. The algorithmic foundations of the framework and its implementation are outlined, and its utility is demonstrated through representative near-surface geological scenarios relevant to site characterization and near-surface geophysics. By facilitating the rapid generation of large-scale model libraries ($N > 10^5$), SoilGen enables comprehensive parametric studies and the training of deep learning inversion networks that require extensive, labeled datasets for shear-wave velocity ($V_S$) profiling and other site characterization tasks.

## 1 INTRODUCTION

Accurate characterization of the near-surface velocity structure is fundamental to seismic site response evaluation, dispersion curve analysis, and a broad range of geotechnical studies. Techniques such as Horizontal-to-Vertical Spectral Ratio (HVSR), Multichannel Analysis of Surface Waves (MASW), and numerical site response modeling all rely on robust subsurface models to yield reliable predictions. However, traditional inversion methods are often computationally intensive and suffer from non-uniqueness, while publicly available borehole datasets containing complete geotechnical properties remain scarce. This data deficit is particularly critical for data-hungry machine learning approaches, which demand hundreds of thousands of labeled models to learn robust mappings from geophysical observations to subsurface properties.

SoilGen addresses this need by programmatically generating one-dimensional layered soil profiles that exhibit realistic thicknesses and velocities, subject to rigorous geophysical constraints. Crucially, the package computes a complete suite of geotechnical parameters—including layer thickness, shear-wave velocity ($V_S$), P-wave velocity ($V_P$), density, and Poisson's ratio—ensuring that each generated model is immediately applicable to dispersion curve forward modeling, HVSR simulation, site response analysis, or machine learning pipelines. Integrated validation routines strictly enforce physical laws, such as ensuring that $V_P$ exceeds $V_S$ and that material properties remain within plausible limits.

The framework facilitates the rapid generation of extensive model libraries ($N > 10^5$), allowing users to assign profiles to predefined geological scenarios, export them in multiple formats, and visualize them via a modern graphical user interface. The remainder of this paper is organized as follows: Section 2 outlines the SoilGen methodology, detailing the scenario definitions, empirical relationships, and implementation specifics. Section 3 presents representative results, illustrating the tool's output through multi-panel figures for various geological settings. Finally, Section 4 concludes with a discussion of the package's broader applications in geotechnical modeling, including its integration with complementary tools such as hvstrip-progressive (Fathizadeh et al., 2025) for advanced layer-stripping analyses.

## 2    METHODOLOGY AND DATA PROCESSING

### 2.1 Profile Generation Algorithm

SoilGen generates randomized 1D soil profiles—typically comprising 3 to 8 layers—by stochastically sampling layer thicknesses and shear-wave velocities ($V_S$), subsequently computing derived elastic properties and validating the physical consistency of each model. The overall generation workflow is illustrated in Figure 1, which depicts the primary interface for parameter definition. To create a synthetic dataset, the user selects a target geological scenario and specifies boundary conditions, including the total number of profiles, layer multiplicity, and permissible depth and velocity ranges.

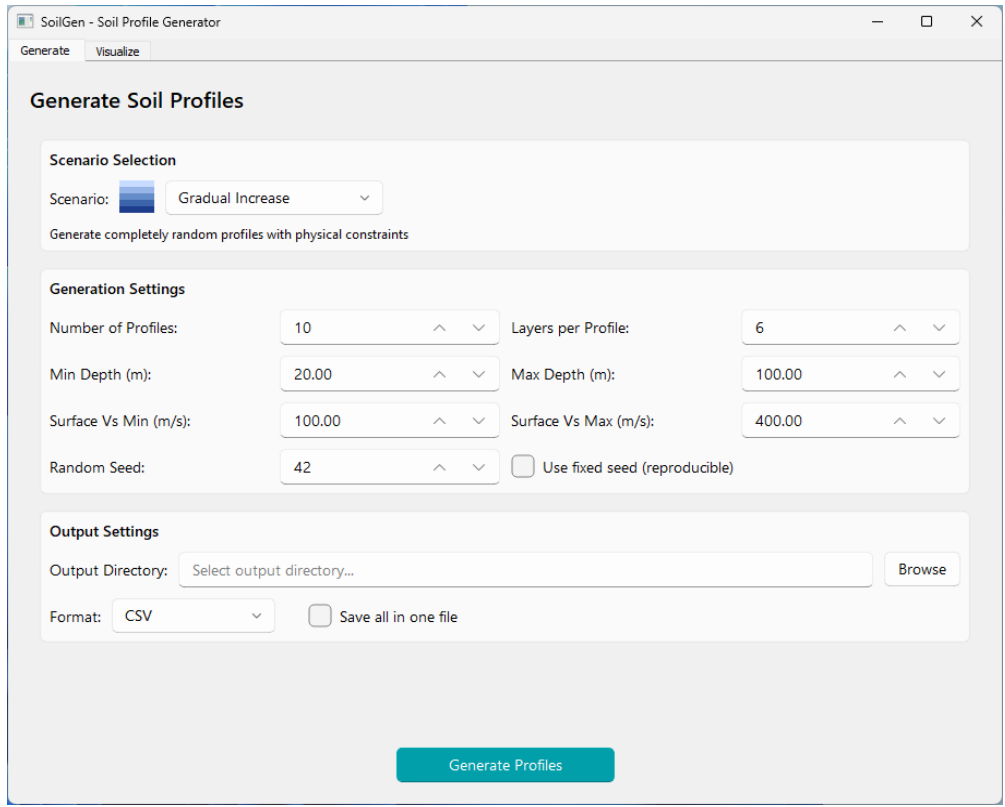

**Figure 1.** The Soil Generation configuration interface. This module serves as the primary control panel for defining stochastic simulation parameters, allowing users to select geological scenarios (e.g., "Gradual Increase"), impose geometric constraints such as depth and surface velocity ranges, and specify the sample size to ensure reproducible dataset generation.

The following generation, the framework provides a dedicated visualization module (Figure 2) for inspecting the resulting model library. This environment facilitates the loading of generated datasets and the interactive plotting of $V_S$, P-wave velocity ($V_P$), and borehole logs. To capture a diverse range of subsurface conditions, the package currently implements six fundamental geological scenarios, whose characteristic velocity signatures are detailed in Table 1. These scenarios are grounded in real-world stratigraphic configurations, such as the sharp soil–bedrock contrasts and complex weathering profiles observed in recent site characterization studies (e.g., Rahimi et al., 2025).

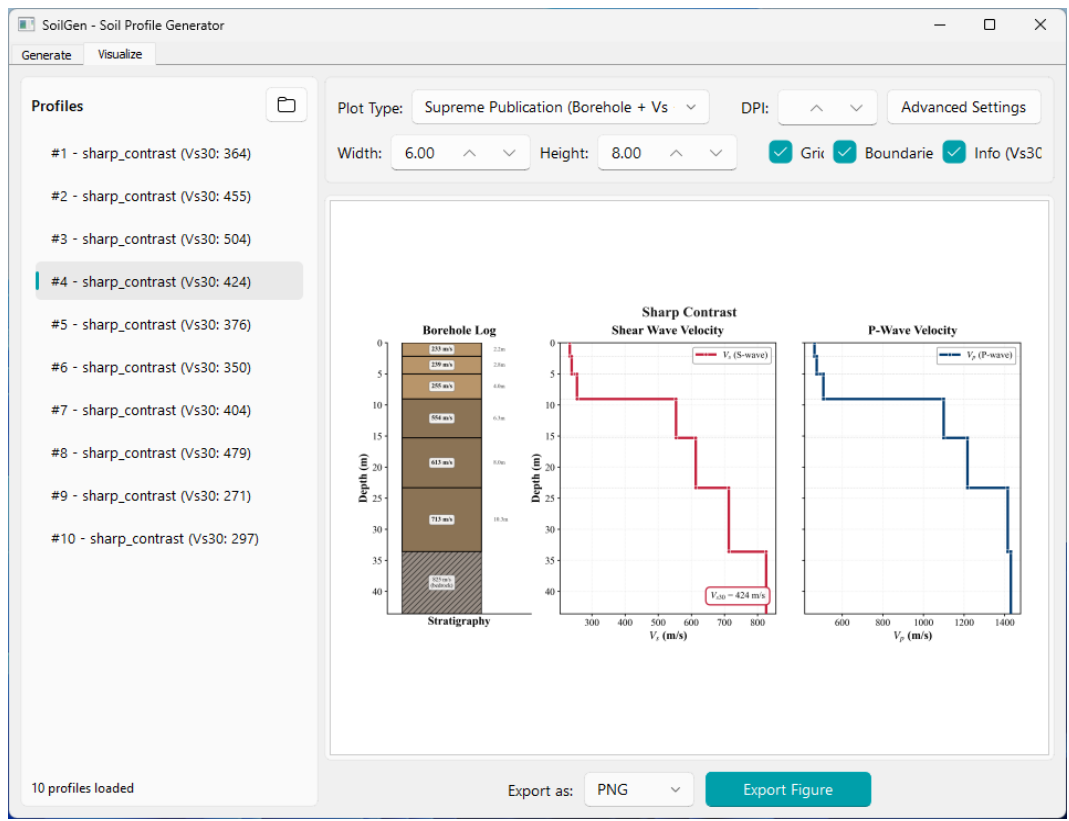

**Figure 2.** The Visualization and inspection interface. The sidebar lists the generated model library, while the main canvas displays a comprehensive multi-panel plot of a selected realization—comprising the lithological borehole log, shear-wave velocity ($V_S$), and P-wave velocity ($V_P$) profiles—to facilitate quality control and graphical export.

Table 1: Geological scenarios implemented in SoilGen and their characteristic velocity profiles.

| Scenario | Physical Setting | Characteristic $V_S$ Pattern |
|---|---|---|
| Gradual Increase | Normal sedimentation sequence typical of alluvial or marine deposits. | $V_S$ increases smoothly with depth, typically exhibiting an increment of 40--80 m/s per layer. |
| Sharp Contrast | Soil--rock interface; specifically, weathered rock overlying competent bedrock. | A distinct, major $V_S$ discontinuity (2--3× increase) at a specific interface, followed by high velocities in deeper strata. |
| Velocity Inversion | A soft geological unit trapped beneath stiffer strata (e.g., peat layer beneath sand). | $CapCapV_S$ generally increases with depth but exhibits a 15--35% reduction within a specific buried interval before recovering. |
| Shallow Bedrock | Thin soil veneer overlying stiff rock; typical of ridge sites or eroded terrains. | A stiff, high-velocity layer ($V_S \approx$ 700--1000 m/s) is present near the surface, overlain by a thin ($<$ 5 m) soil cover. |
| Thick Soft Deposit | Significant thickness of compressible material (e.g., organic clay or peat). | A substantial zone of low $V_S$ ($<$ 200 m/s) persists at intermediate depths (e.g., 20--40 m), bounded by stiffer strata. |

| Thick Stiff Layer | Stiff crust, cemented zone, or consolidated layer embedded within softer sediments. | A thick, high-$V_S$ layer interbedded between softer materials, resulting in a non-monotonic velocity profile. |

## 2.2 Profile generation algorithm

For each realization, the SoilGen algorithm initiates by defining the model domain, specifically determining the number of finite layers and the total soil column thickness within user-specified boundaries. Layer thicknesses are subsequently generated using a stochastic layering-ratio approach: a baseline thickness is multiplied by randomized coefficients to simulate the natural trend of increasing layer thickness with depth. At the same time, a jitter parameter introduces realistic variability to prevent artificial regularity.

Simultaneously, shear-wave velocity ($V_S$) profiles are constructed by drawing a near-surface velocity value from a specified distribution and applying incremental increases for deeper strata. To capture complex geological features, velocity inversions—where $V_S$ temporarily decreases with depth—are introduced based on a user-specified probability and severity ratio. Finally, the deepest finite layer is automatically extended into a half-space by applying a final velocity increment, ensuring a stable boundary condition for wave propagation simulations. For the scenario-based profiles (Table 1), these baseline thickness and velocity distributions are modified to adhere rigidly to the characteristic physical signatures of each geological setting.

Once the primary $V_S$ structure is established, the dependent elastic parameters—P-wave velocity ($V_P$) and density ($\rho$)—are derived using established empirical relationships to ensure physical consistency. One available method assumes constant Poisson's ratios ($\nu$) for soil and rock, respectively, deriving $V_P$ directly from $V_S$ via the fundamental elastic relationship:

$$V_P = V_S \sqrt{\frac{2(1-\nu)}{1-2\nu}}$$

Alternatively, for broader crustal applications, the algorithm employs Brocher's polynomial relation (Brocher, 2005):

$$V_P = 0.9409 + 2.0947 V_S - 0.8206 V_S^2 + 0.2683 V_S^3 - 0.0251 V_S^4$$

Where velocities are in km/s, density is subsequently estimated using Gardner's relation (Gardner et al., 1974):

$$\rho = 0.31 V_P^{0.25} \times 10^3 \; (\text{kg/m}^3)$$

Or via a simplified linear correlation. Poisson's ratio is then back-calculated from the determined $V_P$ and $V_S$ values and rigorously checked to ensure they lie within a physically admissible range (typically $0 < \nu < 0.5$) (Kramer, 1996).

Following parameter assignment, the Soil Profile object computes critical site characterization metrics. The time-averaged shear-wave velocity over the top 30 m ($V_{S30}$) is calculated using the harmonic mean formula:

$$V_{S30} = \frac{30}{\sum_{i=1}^{n} \frac{h_i}{V_{S,i}}}$$

where $h_i$ and $V_{S,i}$ represent the thickness and shear-wave velocity of the $i$-th layer, respectively. Based on standard thresholds, the system automatically classifies each profile according to NEHRP (Building Seismic Safety Council, 2020) and Eurocode 8 (CEN, 2004) standards. Additionally, the fundamental frequency ($f_0$) is estimated via the quarter-wavelength approximation (Kramer, 1996):

$$f_0 \approx \frac{\bar{V}_S}{4H}$$

where $\bar{V}_S$ is the time-averaged shear-wave velocity of the column and $H$ is the total thickness. To ensure robustness, every generated profile undergoes a strict validation protocol. The system verifies that $V_P > V_S$, checks for non-negative values, and confirms that density and thickness fall within realistic geotechnical ranges. Any profile failing these checks is automatically discarded and regenerated.

**2.3 Graphical Interface and Software Architecture**

The SoilGen graphical user interface (GUI) is built upon a modern cross-platform framework, encapsulating the entire generation workflow within a user-friendly environment. The package is designed for broad compatibility, requiring only a standard Python 3.8+ environment and fundamental numerical libraries. Optional dependencies include Matplotlib for advanced plotting and PyQt for GUI execution.

Installation is streamlined via the Python Package Index (pip), which automatically resolves dependencies from the GitHub repository (Fathizadeh, 2025b) and registers entry points for both the command-line interface (CLI) and the GUI. To facilitate reproducible research, generation parameters can be defined via a JSON or YAML configuration file, allowing users to rapidly recreate specific model libraries without modifying the source code. While the GUI provides an accessible entry point for visualization and configuration, all core functionality remains fully accessible via the CLI and Python API for integration into automated high-performance computing pipelines.

**2.4 Interface Design and Visualization**

To facilitate efficient workflow management, SoilGen integrates a graphical user interface (GUI) that abstracts the complexity of the underlying procedural algorithms. Figure 1 delineates the Soil Generation module, the primary control panel where users define the simulation parameters. This interface allows for the selection of specific geological scenarios, the configuration of layer depth and velocity boundaries, and the specification of the total dataset size ($N$). The design prioritizes reproducibility, ensuring that complex generation tasks can be executed and replicated with minimal manual intervention.

Complementing the generation module, the Visualization tab (Figure 2) provides an interactive environment for data inspection. This component lists the generated profile library and offers real-time plotting capabilities for shear-wave velocity ($V_S$), P-wave velocity ($V_P$), and lithological logs. By enabling immediate visual verification of the synthetic models, this tool allows users to assess the geological realism of the stochastic realizations before proceeding to export or downstream analysis. While the GUI offers a streamlined experience for standard applications, the architecture is designed such that all core algorithms remain directly accessible via the command-line interface (CLI) or Python API, supporting integration into larger, automated computational pipelines.

**3 Results**

To demonstrate the capabilities of the SoilGen framework, we generated a comprehensive suite of synthetic profiles corresponding to each of the six geological scenarios defined in Table 1. Representative realizations for these scenarios are presented in Figures 3–8. Each multi-panel figure comprises a lithological log with color-coded layer stiffness, alongside depth-dependent profiles for shear-wave velocity ($V_S$) and P-wave velocity ($V_P$). The lithological log visually conveys the relative thickness and stiffness contrasts of the subsurface layers, while the velocity step plots quantitatively reveal the trends and impedance mismatches critical for wave propagation analysis.

While the examples presented here are generic realizations intended to illustrate the distinct geophysical signatures of each setting, the framework is designed for flexibility. In practice, users can calibrate the stochastic parameters to statistically match regional soil conditions or to generate massive, high-variance libraries ($N > 100{,}000$) suitable for training deep neural networks or conducting extensive parametric sensitivity studies.

**3.1 Gradual Increase**

Figure 3 illustrates a standard depositional environment where $V_S$ increases monotonically from 150 m/s at the surface to 500 m/s in the underlying half-space. The borehole log depicts layers of relatively uniform thickness, while the velocity profiles exhibit a stepwise but smooth gradient. Such profiles serve as robust baseline models for microzonation studies, effectively approximating normally consolidated sedimentary deposits where stiffness is primarily governed by confining stress. As shown in the figure, the steady increase in velocity results in moderate impedance contrasts, minimizing strong resonance peaks typically associated with sharp boundaries.

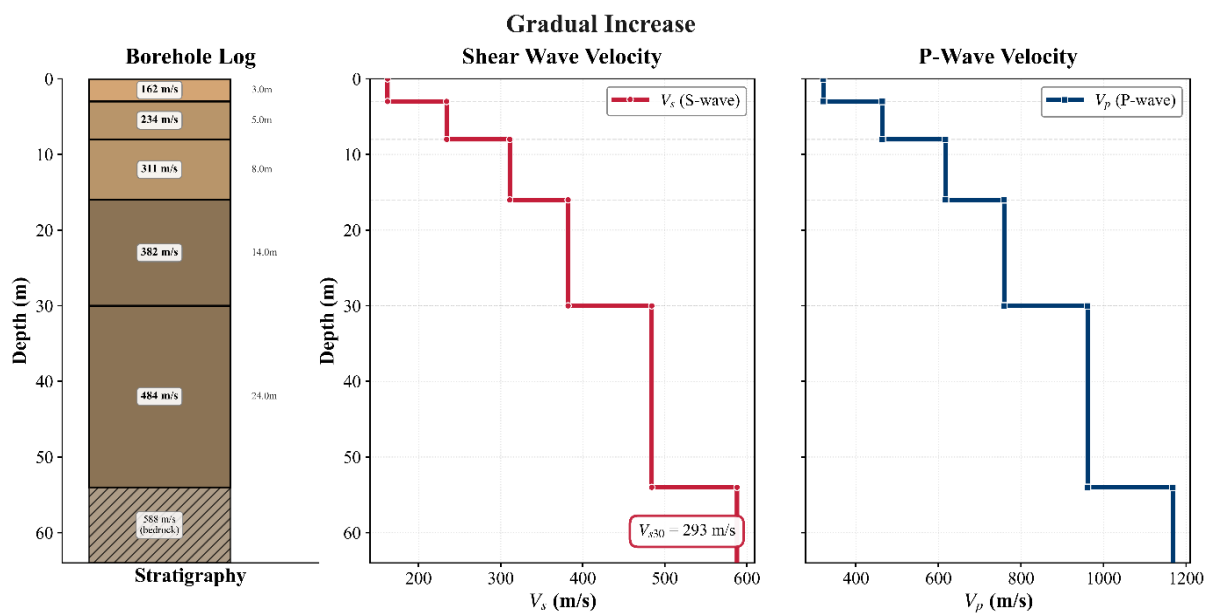

**Figure 3.** Multi-panel figure for the Gradual Increase scenario: borehole log (left), $V_S$ profile (center), and $V_P$ profile (right). $V_S$ values increase steadily, resulting in moderate impedance contrasts.

### 3.2 Sharp Contrast

Figure 4 characterizes a geological setting defined by a pronounced discontinuity in stiffness, typically representing a soil–bedrock interface or a boundary between weathered material and competent rock. In this realization, a significant shear-wave velocity ($V_S$) jump occurs at approximately 10 m depth, marking the transition from a softer overburden to a rigid substratum. As illustrated in the velocity step plots, $V_S$ increases abruptly from 300 m/s to 700 m/s across this interface. This substantial impedance contrast is critical in site response analysis, as it typically yields a high-amplitude fundamental resonance peak in the Horizontal-to-Vertical Spectral Ratio (HVSR) curve, a phenomenon that must be accurately captured in synthetic training datasets.

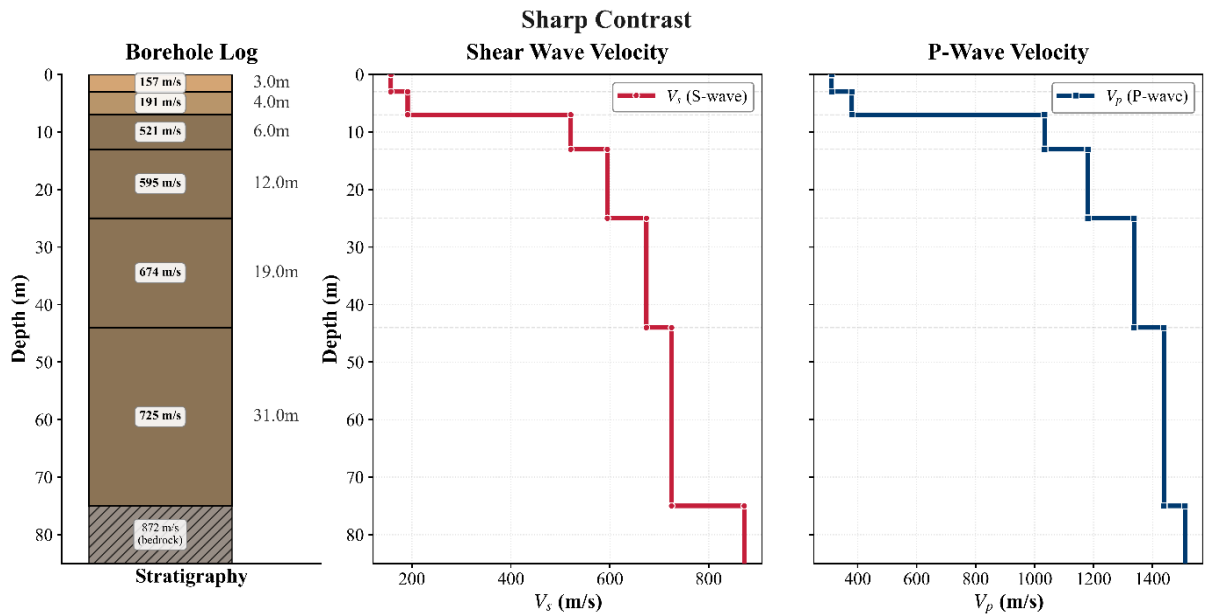

**Figure 4.** Sharp Contrast scenario: A thin soft layer overlies a stiff bedrock, resulting in a substantial jump in both $V_S$ and $V_P$ profiles. This strong impedance contrast is a critical parameter for accurate resonance frequency estimation.

## 3.3 Velocity Inversion

Figure 5 elucidates a more complex stratigraphy characterized by a "velocity inversion," wherein a softer geological unit is interbedded between stiffer strata. In this specific realization, the shear-wave velocity ($V_S$) initially increases from 220 m/s to 260 m/s in the near-surface layers, before exhibiting a significant reduction to 180 m/s within a buried low-velocity zone (LVZ) located at depths of 15–25 m. Following this excursion, the velocity profile recovers, increasing sharply to 800 m/s in the underlying half-space. Such non-monotonic velocity structures are notoriously difficult for traditional inversion algorithms to resolve due to non-uniqueness. Consequently, the ability to stochastically generate these profiles is essential for training robust machine learning models capable of identifying hidden low-velocity layers.

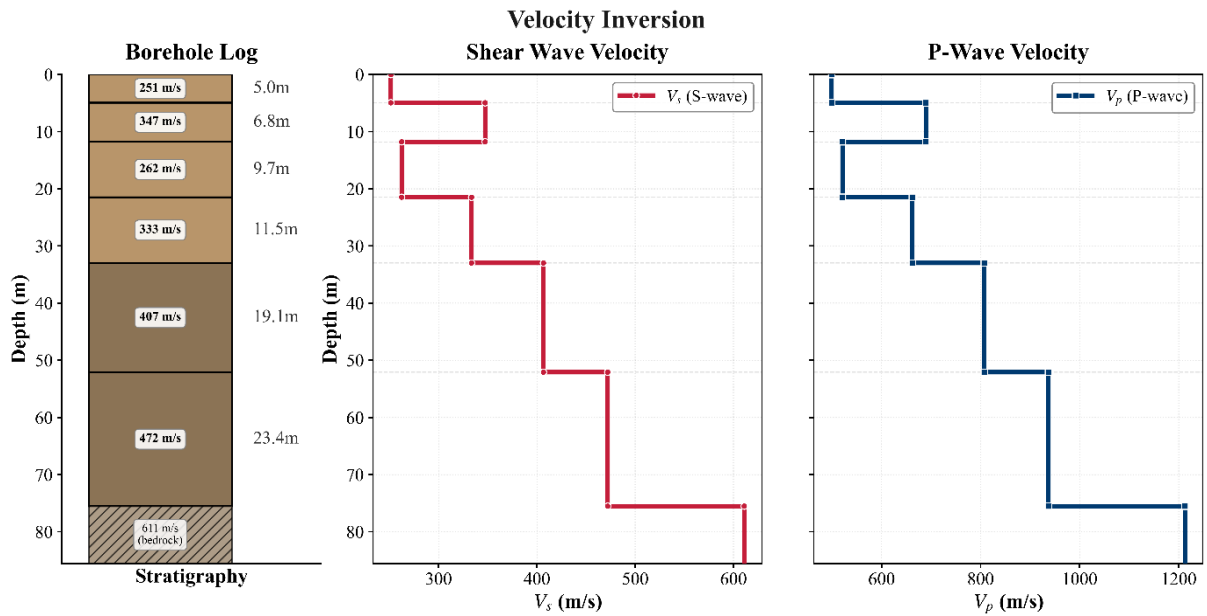

**Figure 5.** Velocity Inversion scenario: The borehole log and $V_S$ profile reveals a low-velocity layer trapped beneath stiffer materials, followed by a recovery in stiffness at depth. The $V_P$ profile mirrors this trend, ensuring physical consistency across elastic parameters.

**3.4 Shallow Bedrock**

Figure 6 delineates a "Shallow Bedrock" scenario, characterized by a thin overburden of soft soil overlying a stiff rock mass. In this realization, the shear-wave velocity ($V_S$) increases rapidly to exceed 900 m/s within the top 7 m of the profile. Such velocity structures are geologically representative of ridge sites, glacially scoured terrains, or areas with minimal sediment accumulation. Consequently, the rapid stiffness increase at shallow depths yields high fundamental resonance frequencies ($f_0$), a critical parameter in seismic hazard assessment for stiff sites.

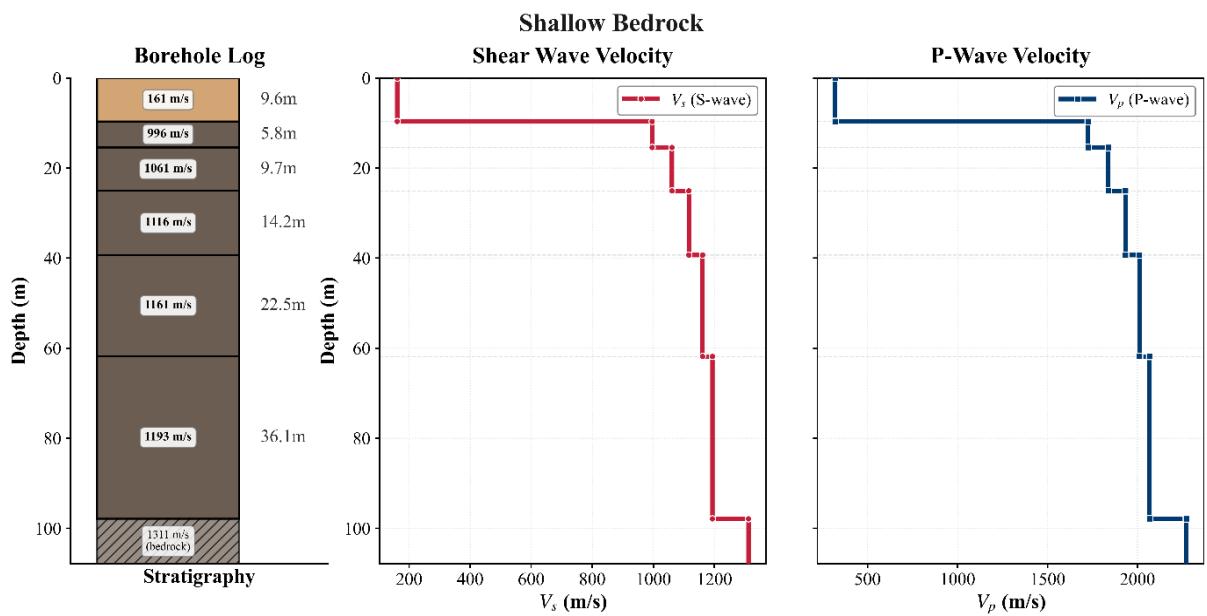

**Figure 6.** Shallow Bedrock scenario: A thin, low-velocity layer overlies a rigid bedrock with $V_S$ exceeding 900 m/s. The rapid increase in stiffness results in a high-frequency fundamental resonance mode, distinguishing this setting from deeper sedimentary basins.

### 3.5 Thick Soft Deposit

Figure 7 characterizes a "Thick Soft Deposit" scenario, defined by a substantial interval of low-stiffness material (e.g., organic clay or peat) extending between depths of 20 m and 40 m. Throughout this specific zone, the shear-wave velocity ($V_S$) remains consistently low (≈ 200–220 m/s), effectively trapping seismic energy before the profile eventually recovers to 800 m/s in the underlying half-space. Such stratigraphic configurations are critically important in earthquake engineering, as they often result in significant ground motion amplification and long-period resonance. These phenomena must be adequately represented in synthetic training sets for hazard analysis.

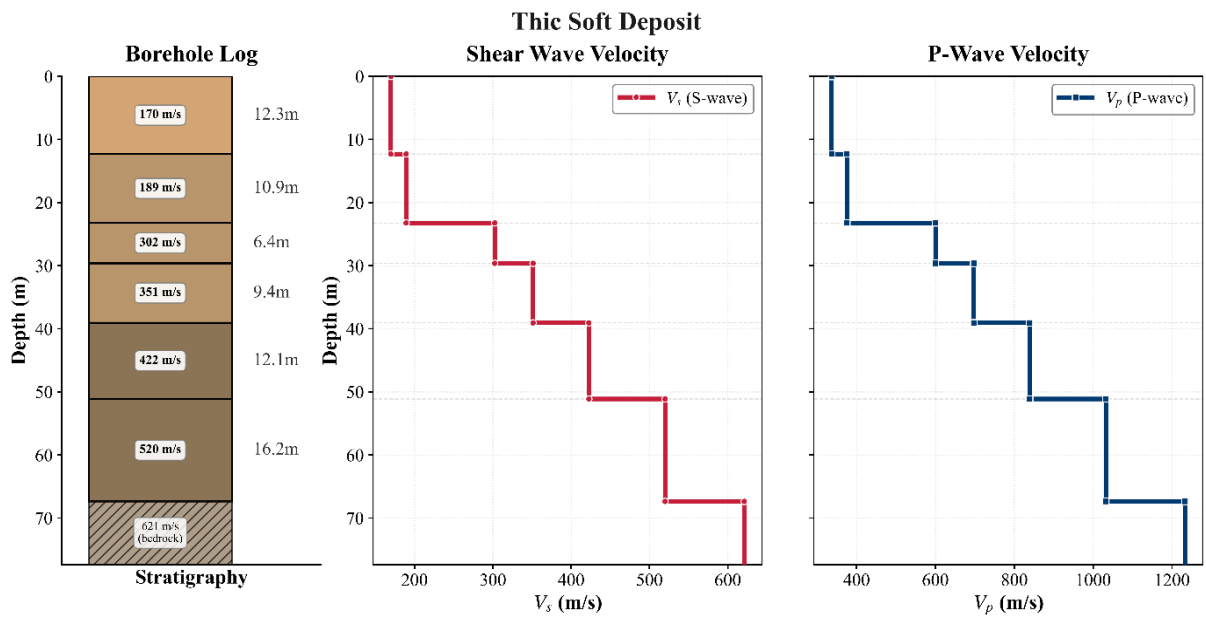

**Figure 7.** Thick Soft Deposit scenario: The borehole log reveals a substantial low-velocity zone at intermediate depths. The $V_S$ and $V_P$ profiles remain depressed within this layer before rising sharply at the base interface, creating a waveguide effect for trapped energy.

### 3.6 Thick Stiff Layer

Figure 8 portrays a complex stratigraphy featuring a "Thick Stiff Layer" interbedded within softer sedimentary deposits. In this realization, $V_S$ increases from 200 m/s to 700 m/s within a consolidated intermediate zone (e.g., a calcrete layer or stiff crust), before decreasing slightly in the underlying softer stratum and ultimately rising again at depth. This non-monotonic velocity structure generates complex impedance profiles that can result in higher-mode surface wave propagation and de-amplification at specific frequencies. Consequently, the ability to model such reversals is essential for testing the resolution limits of inversion algorithms, which often struggle to resolve stiff inclusions within softer media.

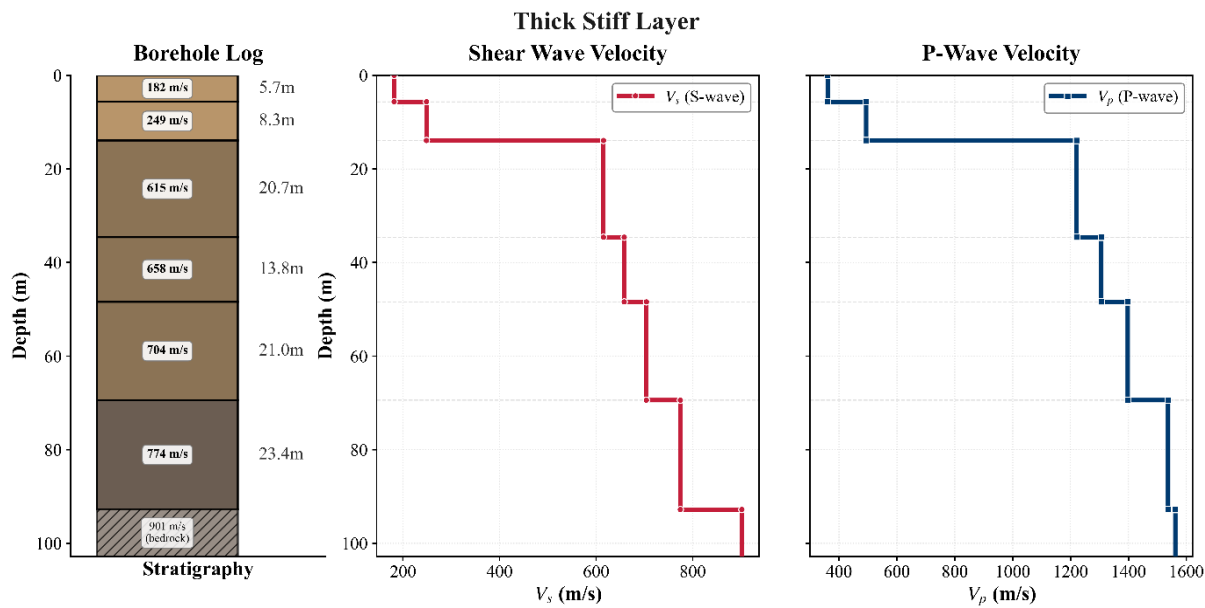

**Figure 8.** Thick Stiff Layer scenario: An intermediate stiff layer produces a non-monotonic velocity profile with multiple strong impedance contrasts. This setting poses significant challenges for conventional inversion techniques and has distinct implications for wave propagation and resonance.

## 4    Conclusion

SoilGen establishes a systematic, open-source framework for the procedural generation of synthetic soil profiles, rigorously computing a full suite of geotechnical properties—including layer thickness, shear-wave velocity ($V_S$), P-wave velocity ($V_P$), density, and Poisson's ratio—subject to physical constraints. By implementing six geologically motivated scenarios, the package effectively captures a diverse range of subsurface conditions, from gradual stiffness gradients and sharp soil–rock interfaces to complex velocity inversions and deep soft deposits. The framework's capacity to rapidly generate extensive, parameter-configurable datasets ($N > 10^5$), coupled with its flexible export options and intuitive graphical interface, positions it as a critical utility for dispersion curve modeling, HVSR forward simulation, site response analysis, and data-driven machine learning research.

The representative results presented herein demonstrate that each scenario yields distinct impedance contrasts and velocity signatures, which fundamentally control resonance frequencies and wave propagation characteristics. Consequently, researchers can leverage SoilGen to construct robust training sets for deep learning models tasked with inferring velocity structure from HVSR or dispersion observations. Furthermore, the seamless integration with standard formats (e.g., Geopsy) facilitates direct application in microzonation studies. When employed in conjunction with complementary tools such as hvstrip-progressive(Fathizadeh, 2025a; Fathizadeh et al., 2025), SoilGen enables comprehensive progressive layer-stripping analyses, allowing researchers to causally link resonance peaks to specific stratigraphic impedance contrasts. By simulating the complete elastic soil column rather than isolated velocity parameters, the package ensures compatibility across a broad spectrum of geotechnical analyses.

Looking ahead, future development will focus on incorporating data-driven parameter distributions derived from regional borehole databases, introducing lateral variability for 2D/3D modeling, and expanding export capabilities to finite-element solvers. Ultimately, SoilGen serves as a robust foundation for the reproducible, stochastic generation of synthetic geotechnical models, bridging the gap between theoretical formulation and data-intensive computational geophysics.

Rahimi, M., Wood, C. M., Fathizadeh, M., & Rahimi, S. (2025). A multi-method geophysical approach for complex shallow landslide characterization. *Annals of Geophysics*, *68*(3), NS336. https://doi.org/10.4401/ag-9203